\documentclass[lettersize,journal]{IEEEtran}
\usepackage{amsmath,amsfonts}
\usepackage{algorithmic}
\usepackage{algorithm}
\usepackage{array}
\usepackage[caption=false,font=normalsize,labelfont=sf,textfont=sf]{subfig}
\usepackage{textcomp}
\usepackage{stfloats}
\usepackage{url}
\usepackage{verbatim}
\usepackage{graphicx}
\usepackage{cite}
\begin{document}

\title{Enhanced Index Modulation Aided Non-Orthogonal Multiple Access via Constellation Rotation}%{Enhanced Index Modulation Aided Non-Orthogonal Multiple Access with the Rotation of Signal Constellation of Low Power Users}

\author{Ronglan Huang, Fei ji, Zeng Hu, Dehuan Wan, Pengcheng Xu, and Yun Liu
        % <-this % stops a space
%\thanks{This paper was produced by the IEEE Publication Technology Group. They are in Piscataway, NJ.}% <-this % stops a space
%\thanks{Manuscript received April 19, 2021; revised August 16, 2021.}
\thanks{R. Huang and F. Ji are with School of Electronic and Information Engineering, South China University of Technology, Guangzhou, 510641, China (e-mail: eerlhuang@mail.scut.edu.cn, eefeiji@scut.edu.cn).}
\thanks{Z. Hu is with College of Information Science and Technology, Zhongkai University of Agriculture and Engineering, Guangzhou 510225, China, and also with Guangdong Provincial Key Laboratory of Short-Range Wireless Detection and Communication, South China University of Technology, Guangzhou 510640, China (e-mail: huzeng@zhku.edu.cn).}
\thanks{D. Wan is with School of Internet Finance and Information Engineering, Guangdong University of Finance, Guangzhou 510521, China, and also with Guangdong Provincial Key Laboratory of Short-Range Wireless Detection and Communication, South China University of Technology, Guangzhou 510640, China (e-mail: wan\_e@gduf.edu.cn).}
\thanks{P. Xu is with School of Financial Mathematics and Statistics, Guangdong University of Finance, Guangzhou 510521, China (e-mail: 47-105@gduf.edu.cn).}
\thanks{Y. Liu is with School of Internet Finance and Information Engineering, Guangdong University of Finance, Guangzhou 510521, China (e-mail: yunliu@gduf.edu.cn).}
\thanks{%This work was supported in part by the National Natural Science Foundation of China under grant number 61971149, in part by the Special Projects in Key Fields for General Universities of Guangdong Province under Grant 2020ZDZX3025, in part by the Guangdong Basic and Applied Basic Research Foundation under Grant 2021A1515011657. 
%Corresponding author: \textit{Z. Hu (huzeng@zhku.edu.cn) and D. Wan (wan\_e@gduf.edu.cn)}
}
}
\markboth{Journal of \LaTeX\ Class Files,~Vol.~14, No.~8, August~2023}%
{Shell \MakeLowercase{\textit{et al.}}: A Sample Article Using IEEEtran.cls for IEEE Journals}


\maketitle

\begin{abstract}
Non-orthogonal multiple access (NOMA) has been widely nominated as an emerging spectral efficiency (SE) multiple access technique for the next generation of wireless communication network. To meet the growing demands in massive connectivity and huge data in transmission, a novel index modulation aided NOMA with the rotation of signal constellation of low power users (IM-NOMA-RC) is developed to the downlink transmission. In the proposed IM-NOMA-RC system, the users are classified into far-user group and near-user group according to their channel conditions, where the rotation constellation based IM operation is performed only on the users who belong to the near-user group that are allocated lower power compared with the far ones to transmit extra information. In the proposed IM-NOMA-RC, all the subcarriers are activated to transmit information to multiple users to achieve higher SE. With the aid of the multiple dimension modulation in IM-NOMA-RC, more users can be supported over an orthogonal resource block. Then, both maximum likelihood (ML) detector and successive interference cancellation (SIC) detector are studied for all the user. Numerical simulation results of the proposed IM-NOMA-RC scheme are investigate for the ML detector and the SIC detector for each users, which shows that proposed scheme can outperform conventional NOMA.
\end{abstract}


\section{Introduction}
Non-orthogonal multiple access (NOMA) is regarded as one of key technologies of the sixth generation (6G) wireless communication networks \cite{Chowdhury20206GWC,Ding2017NOMA,Yang20196G,Wan2018NOMASurvey}. In conventional orthogonal multiple access (OMA) systems, such as time division multiple access (TDMA) and orthogonal frequency division multiple access (OFDMA), which can only serve one user in each orthogonal resource block \cite{LDai2015NOMA5G,YueX2024RISNONA,WanD2018CNOMA}. As the rapid development of Internet of Things (IoT) in industry, smart agriculture and hospital, the devices that connect to the wireless networks increase exponentially in recent years. How to support such huge devices become an urgent problem for both academic and industry. 
In NOMA, the base station (BS) can serve more than one users by utilizing the same channel to achieve higher spectral efficiency (SE). 
In power domain NOMA (PD-NOMA), multiple symbols are  transmitted simultaneously in the same time slot or subcarrier for different users by allocating different power according their distance or the quality of the experienced channel in downlink transmission \cite{PeiX2022NOMAPLM, LiuX2017NOMARelay, ChenS2017PDMA}. and the dynamic power allocation scheme is proposed

Currently, NOMA has been used in many application scenarios, such as downlink transmission, uplink transmission, and cooperative communication \cite{LDai2018NOMASurvey}. For the NOMA-based cooperative relay network that formed by downlink and uplink transmissions, for example, its achievable sum-rates are analyzed with the assumptions of both the statistical and instantaneous channel state information are known at the transmitter \cite{Wan2019SumRate}. Besides, the performance of NOMA can be further enhanced by combining it with the reconfigurable intelligent surface (RIS) technology \cite{HouT2020RISNOMA, HouT2024RISNOMA}. More specifically, a RIS-aided NOMA scheme is proposed for wireless communication networks, in which one can see that the proposed scheme shows its superiority over the existing schemes in the low and high signal-to-noise-ratio regimes \cite{HouT2023PerfNOMAR}. To reduce the detection complexity, on the other hand, constellation rotation method is introduced into NOMA system, and a specific case with the rotation angel optimization for constellation is studied in \cite{ZhangJ2016RCNOMA}. 

In PD-NOMA, the near users not only need to detect its own symbols, but also need to detect the far users' symbols due to the superimposed symbol transmitted by the transmitter. Specifically, the transmitted superposition symbol, who formed by muiti-user's intended symbols with different power, is transmitted from the BS to all receivers simultaneously. Typically, higher power is allocated to the far user due to its poor channel condition, while lower power is assigned for the near user due to its good channel quality. For each user, the successive interference cancellation (SIC) detector detects the symbols of the far users first and cancels the obtained symbols from the received signals, and repeats the detection processes until its intended symbol is acquired \cite{Islam2017PDNOMA,A2020SICIMNOMA}.

More recently, index modulation (IM) is proposed as a promising digital modulation technique for next generation wireless communication networks due to its better bit error rate (BER) performance and flexible transceiver designs\cite{LiJ2022IM6G,Wen2021IMBook,Basar2016mag,Wen2021CyclicDelay}. In IM aided transmission schemes, the information bits are conveyed in two ways, where the first way is conventional modulated symbols and the second is the index patterns of communication resource blocks \cite{Wen2017IMBook,basar2017index,LiJ2019Layered}. In MIMO system, generalized spatial modulation (GSM) selects multiple transmit antennas to send modulated symbols, in which the index patterns of the selected transmit antennas are also transmit information bits implicitly \cite{Yang2015SM,Wen2019SurveySM,LiJ2017PSM}. To further optimize the transmit antennas selection, an enhanced spatial modulation with generalized antenna selection with variable of transmit antennas is proposed \cite{QingH2021Enhancedsm}.

By utilizing the index resource provided in orthogonal frequency division multiple (OFDM), OFDM with IM (OFDM-IM) is proposed as an improvement of conventional OFDM, in which the available subcarriers are divided into multiple subblocks to perform independent IM operation \cite{LiJ2020IMAccess,Basar2013IM}. Then, the upper bound of the BER performance for OFDM-IM over Rayleigh fading channel is analyzed theoretically by assuming the optimal maximum likelihood (ML) detector is employed \cite{Basar2013IM,LiQ2020Principles}.  Owing to the robustness of the activated subcarrier patterns (ASPs), OFDM-IM has potential to achieve better BER performance compared with conventional OFDM in low-to-middle SE \cite{Wen2016Achievable,Li2018OFDMspread,Li2020Cognitive}.

To avoid the exponential increasing of the computational complexity in the ML criterion based detection due to the dependence of the active states of the subcarriers within a subblock, a serial of low complexity detectors are proposed for OFDM-IM. In log-likelihood ratio (LLR) detector, LLR value of each subcarrier, which indicates the active state of subcarrier, is calculated to obtained the estimate of ASP in the first stage. Then, the modulated symbols are detected according to the obtained ASP in the second stage \cite{Zheng2015IQIM,Wen2017Enhanced}. Since the error propagation that a mistake of ASP leads to error modulated symbol vector, LLR detector suffer from a minor BER performance loss compared with the optimal ML detector. By taking advantage of the orthogonality of subcarriers in the subblock, a subcarrier-wise ML detector, which can achieve the same BER performance as the brute force ML detector that implemented in subblock-wise with significantly low computational complexity, is proposed for OFDM-IM \cite{Hu2022CIM}. In subcarrier-wise ML detector, we first calculate the metrics of each subcarrier under the condition of active and inactive states, and then get the estimate of the subblock  based on the metrics of each subcarrier in subblock-wise.

Owing to the advantages of IM, the index domain in NOMA system is also exploited to further improve the system performance. In \cite{Alm2021IMNOMA}, a novel IM based NOMA downlink scheme is proposed, which serves the users based on IM concept within one subblock and determines the power levels of each users over an orthogonal resource block based on NOMA concept. In MIMO system, an IM aided NOMA transmission technique is proposed which can support massive connectivity in a wireless network with better sum rate and BER performance \cite{Kum2020DoNOMA}. SM aided cooperative NOMA three-node cooperative relaying system is proposed, which can improve power efficiency in transmission \cite{LiQ2019SMNOMA}. Then, quadrature spatial modulation based NOMA scheme for next generation vehicular networks is proposed in \cite{LiJ2021GQSM}.  By calculating the LLR values of each subcarriers, a two-stage LLR detector is proposed, which finds the activated  subcarriers based on the LLR values and detects the modulated symbols corresponding to the subcarriers obtained in the first stage \cite{Sabud2022LLRNOMA}.

As seen in the current related works, the introducing of IM into NOMA has potential to further improve the outage probability as well as system BER performance of conventional NOMA. In the above IM aided NOMA systems, the IM operation is performed  by utilizing the active states of the subcarriers within a subblock, which suffers a minor SE loss due to the inactive subcarriers. Motivated by the distinguishable constellations generated by rotating the conventional constellation, a rotation constellations based IM scheme can be introduced into NOMA to increase the sum rate.  
In this paper, we propose a novel IM-NOMA with rotation constellation (IM-NOMA-RC) scheme in which only the near devices are selected to perform IM operation is proposed.
In the proposed IM-NOMA-RC, the users are divided into far user group and near user group according to their distance to the BS. Since all the subcarriers are activated to transmit modulated symbols, higher SE can be achieved in proposed IM-NOMA-RC scheme. The main contributions of this paper are summarized as follows.

\begin{itemize}
	\item A novel IM-NOMA-RC scheme is proposed for downlink transmission, in which the IM operation are performed on the near users within an orthogonal resource block. In IM-NOMA-RC, the IM operation is performed by rotating constellation for the near users. Since all the subcarriers are activated to transmit modulated symbols, IM-NOMA-RC can achieve higher sum rate due to extra information bits are conveyed by IM for the near users. 
		
	\item 
	In IM-NOMA-RC, the rotation of constellations are utilized to performed IM operation for near users within each orthogonal resource block. 
	Since conventional modulated symbols are transmitted by the far users, the BER performance of far users maintain same as conventional NOMA. An union bound on the BER performance is derived in closed form by assuming the ML detector is employed. 	
	 
	\item To reduce burden of computational complexity of IM-NOMA-RC, a joint SIC and LLR detector is proposed to the IM-NOMA-RC system. Numerical simulation results demonstrate that the proposed IM-NOMA-RC has potential to achieve superiority system performance over conventional NOMA system. 
\end{itemize}

The remainder of this paper is organized as follows. Section \ref{System-model} introduces the system model of IM-NOMA-RC and the transmitted signal model in IM operation. The detectors for IM-NOMA-RC system is proposed and the error performance is investigated in Section \ref{Detectors}. The computer simulation and numerical results are provided in Section \ref{Simulation-results}. Finally, the paper is concluded in Section \ref{Conclusions}.

\textit{Notation:} $ \mathbf{X} $ denotes a matrix and $ \mathbf{x} $ denotes a column-vector.
${\left(  \cdot  \right)^T}$ and ${\left(  \cdot  \right)^H}$ denote transposition and Hermitian transposition of a matrix or a vector, respectively. 
$\textrm{diag}\left\{ \mathbf{x} \right\}$ returns a diagonal matrix whose diagonal elements are included in $\mathbf{x}$.
%$ \mathbf{I_{k}} $ denotes the $ k \times k $ identity matrix. 
$x \sim \mathcal{CN}\left( {0,\sigma _x^2} \right)$ represents the distribution of a zero mean circularly symmetric complex Gaussian random variable $x$ with variance ${\sigma _x^2}$. 
$\left\lfloor  \cdot  \right\rfloor $ is the integer floor operation and $\emptyset $ denotes the empty set.
$\lVert \cdot \rVert $ stands for Frobenius norm. 
%$\left| \cdot \right|$ denotes the absolute value of a complex number.
$Pr\left( \cdot \right) $ denotes the probability of an event. 
$ \mathcal{S}$ denotes the complex symbol constellation of size $\bar{M}$.
${\rm{FFT}}\left\{  \cdot  \right\} $ denotes the fast Fourier transform (FFT) operator, and ${\rm{IFFT}}\left\{  \cdot  \right\} $ denotes the inverse FFT operator.

\begin{figure}[!t]
	\centering
	\includegraphics[width=3.0in]{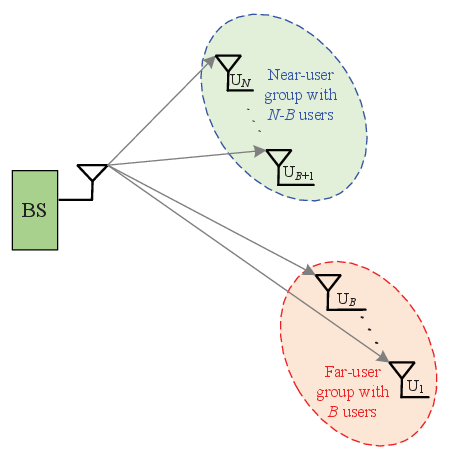}
	\caption{The basic system model of IM-NOMA-RC.}
	\label{fig_1_IM_NOMA_RC}
\end{figure}

\section{proposed IM-NOMA-RC System Model}\label{System-model}
In this section, a downlink scenario is considered, where $N$ users or devices that equipped with single receive antenna are served by a BS with different distances. In the proposed IM-NOMA-RC scheme, all the devices are divided into two subgroups to perform IM operations, where the basic system model of the proposed scheme is illustrated in Fig. \ref{fig_1_IM_NOMA_RC}, where a total of $N$ users are classified into two groups: far-user group and near-user group. More specifically, there are $ B $ cell-edge users with high power level in the far-user group and $N-B$ cell-center users with low power level in the near-user group, respectively. In IM-NOMA-RC system, the available bandwidth is divided into $L $ orthogonal subcarriers as that in OFDMA system, in which each subcarrier is designed to serve a set of devices in downlink transmission. In what following, and without loss of generality, the signal processing of the proposed algorithm is analyzed based on a signal chunk. 

In downlink IM-NOMA-RC system, the signals to be sent to different users are also multiplexed in power domain over the orthogonal subcarriers with different power levels.
Similar to OFDM-IM system, the input information bits for the $ k $th subcarrier are partitioned into two parts, where the first part $p_1=N\log _2M$ symbol bits are send into an $ M $-ary modulator to generate a modulated symbol vector for $ N $ users over the $ k $th subcarrier, and the second $p_2=\lfloor \log _2\left( N-B+1 \right) \rfloor $ bits are denoted as index bits to select an assigned rotation angel for IM operation of the near users.
According to the input $ p_1 $ symbol bits, the normalized symbols for both far and near users can be expressed as $\mathbf{s}_k=\left[ s_1,\dots ,s_B,s_{B+1},\dots ,s_N \right] ^T$. Let us define that the distance of user $ U_{n} $ from the BS is given by $ d_{n} $, where $d_1>d_2>\cdots >d_N$. 
In IM-NOMA-RC, the allocated power coefficient of the $ n $th user is defined as $\alpha _n$, where the power coefficients of all the users satisfy $\alpha _1>\alpha _2>\cdots >\alpha _N$ and $\sum_{n=1}^N{\alpha _n}=1$. The basic transmitted symbol model of NOMA from the BS in downlink transmission is given by 
\begin{equation}\label{Basic-NOMA-x}
	x_{k}=\sum_{n=1}^B{\sqrt{\alpha _nP_T}}s_n+\sum_{n=B+1}^N{\sqrt{\alpha _nP_T}}s_n,
\end{equation}
where $ s_n $ denotes $ n $th element of transmitted symbol vector $ \mathbf{s}_k $ that corresponding to the $ n $th user, $P_T$ is the total transmit power. 
In the proposed IM-NOMA-RC, the IM operation is performed only on the transmitted symbols for near users, which are given by $ s_{i} $ with $i=B+\text{1,}\dots ,N$.  A superimposed symbol after the constellation rotation based IM operation according to the input $ p_2 $ index bits can be expressed as
\begin{equation}\label{IM-NOMA-x-Model}
	\begin{aligned}
	x_{k}^{\varphi}= &\sum_{n=1}^B{\sqrt{\alpha _nP_T}}s_n+\sum_{n=B+1}^{N-\varphi}{\sqrt{\alpha _nP_T}}s_n
	\\
	& +e^{j\frac{\pi}{2}}\left( \sum_{n=N-\varphi +1}^N{\sqrt{\alpha _nP_T}}s_n \right)  
\end{aligned} , 
\end{equation}
where $ \varphi $ is the decimal number of the index bits, $\sum_{n=J}^L{\left( \cdot \right)}=\emptyset $ if $L<J$. To illustrate the constellation rotation based IM operation more clearly, Table \ref{IM-mapping-table} provide a generalized lookup table for the mapping relationship. As shown in the table, the IM operator selects a portion of symbols of the near users to perform rotation operation. Note that, the output of the rotation constellation based IM is conveyed implicitly by the transmitted symbols corresponding to the near users. Then, the superimposed symbols in frequency domain are processed by inverse fast Fourier transform (IFFT) operation to get the time domain signals. After this point, the same procedures of the conventional OFDM are performed to get the final transmitted signal, which are transmit to the multiple users by the BS. 

\begin{table*}[!t]
	\caption{A generalized example of lookup table for IM operation for near users. \label{IM-mapping-table}}
	\centering
	\begin{tabular}{|c||c|}
		\hline
		Index bits $ p_2 $  & Superimposed symbol $ x_{k}^{\varphi} $  \\
		\hline
		$\left[ 0\cdots {0\:}0 \right] $ & $x_{k}^{0}=\sum_{n=1}^B{\sqrt{\alpha _nP_T}}s_n+\sum_{n=B+1}^N{\sqrt{\alpha _nP_T}}s_n$  \\
		\hline	
	$\left[ 0\cdots {0\:}1 \right] $ & $x_{k}^{1}=\sum_{n=1}^B{\sqrt{\alpha _nP_T}}s_n+\sum_{n=B+1}^{N-1}{\sqrt{\alpha _nP_T}}s_n+e^{j\frac{\pi}{2}}\left( \sum_{n=N}^N{\sqrt{\alpha _nP_T}}s_n \right) $  \\
	\hline
	$\left[ 0\cdots \text{1\:}0 \right] $ & $x_{k}^{2}=\sum_{n=1}^B{\sqrt{\alpha _nP_T}}s_n+\sum_{n=B+1}^{N-2}{\sqrt{\alpha _nP_T}}s_n+e^{j\frac{\pi}{2}}\left( \sum_{n=N-1}^N{\sqrt{\alpha _nP_T}}s_n \right) $  \\
	\hline	
	$\vdots $  & $\vdots $  \\	
	\hline
	$\left[ 1\cdots {1\:}1 \right] $ & $x_{k}^{N-B}=\sum_{n=1}^B{\sqrt{\alpha _nP_T}}s_n+e^{j\frac{\pi}{2}}\left( \sum_{n=B+1}^N{\sqrt{\alpha _nP_T}}s_n \right) $  \\
	\hline	
	\end{tabular}
\end{table*}

At the receiver side, the time domain received signal are processed similar as the conventional OFDM system first and a fast Fourier transform (FFT) operation is performed to obtain the frequency domain received signals. 
%Without loss of generality, we focus on the signal processes of the superimposed symbols over the $ k $th subcarrier and omit the subscript $ k $ for brevity.
For the $ n $th user at the receiver side, the superimposed frequency domain received signal at the $ k $th subcarrier can be expressed as
\begin{equation}\label{y-signal-n}
	y_{n,k}=h_{n,k}x_{k}+w_{n,k} ,
\end{equation}
where $ h_{n,k} $ denotes the channel frequency response between the BS and the $ n $th user, $ w_{n,k} $ denotes the zero-mean additive white Gaussian noise (AWGN) with $\sigma ^2$ variance. We assume that the signals experience independent and identically distributed frequency selective Rayleigh fading channel from BS to different users, which are known by receivers but not available at the BS. The SE of the IM-NOMA-RC system can be expressed as
\begin{equation}\label{SE-per-Subcarrier}
	\phi =N\log _2M+\lfloor \log _2\left( N-B+1 \right) \rfloor ,
\end{equation}
where the first part is symbol bits transmitted by conventional modulated symbols, and the second part is index bits convey by the rotation angel of the symbols of the near users.

\begin{figure}[!t]
	\centering
	\includegraphics[width=2.5in]{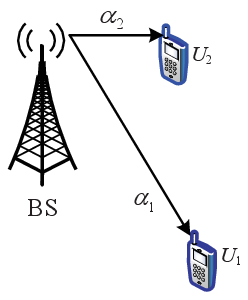}
	\caption{IM-NOMA-RC system with a two-user case.}
	\label{fig_2_IM_NOMA-2users}
\end{figure}

To illustrate the proposed IM-NOMA-RC system more clearly, a realization of IM-NOMA-RC with parameters $N=2$, $B=1$ and BPSK constellation is given as follows. 

\textit{Example:} For the IM-NOMA-RC system with two users served by a BS as illustrated in Fig. \ref{fig_2_IM_NOMA-2users}, where $ U_{1} $ and $ U_{2} $ are corresponding to the far and near user groups, respectively. 
In the IM operation, only the transmitted symbols for near user are rotated according to the index bits to transmit more information bits implicitly. In this case, every subcarrier can transmit $p_2=1$ index bits, where the mapping rule is given by $\left[ 0 \right] \leftrightarrow x_{k}^{0}$ and $\left[ 1 \right] \leftrightarrow x_{k}^{1}$, respectively. According to IM principle given above, the transmitted symbol model over the $k$th subcarrier can be expressed as
\begin{equation}\label{Mapping-rule}
	\begin{cases}
		x_{k}^{0}=\sqrt{\alpha _1P_T}s_1+\sqrt{\alpha _2nP_T}s_2		\\
		x_{k}^{1}=\sqrt{\alpha _1P_T}s_1+\left( \sqrt{\alpha _2P_T}s_2 \right) e^{j\frac{\pi}{2}}		\\
	\end{cases},
\end{equation}
where $ s_1 $ is the transmitted symbol of the far user, and $ s_2 $ the transmitted symbol of the near user. 

\begin{figure*}[!t]
	\centering
	\includegraphics[width=6.0in]{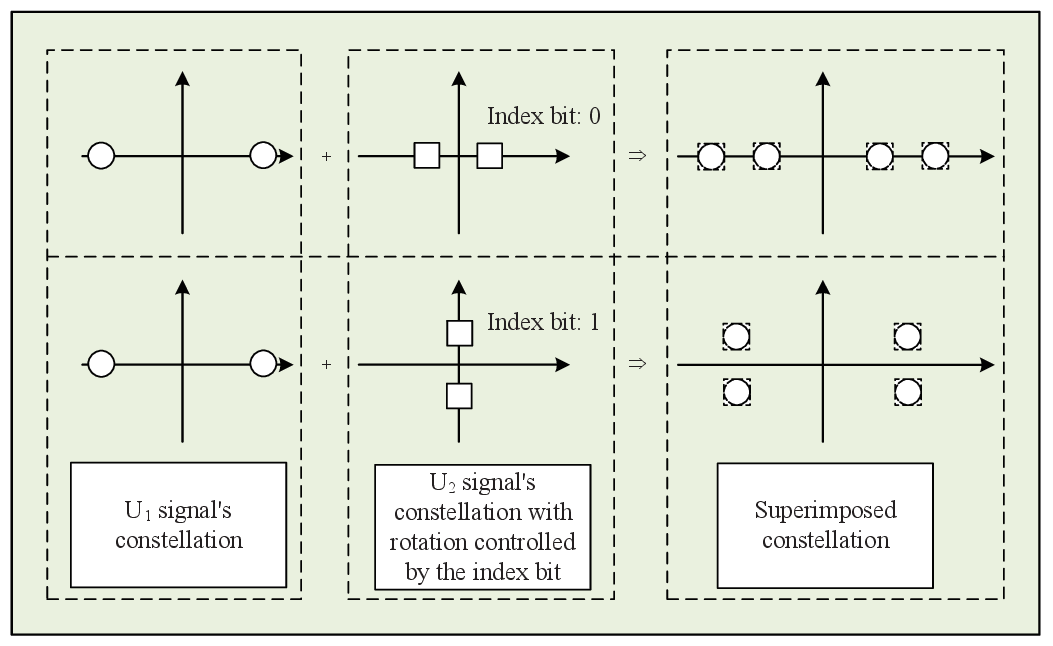}
	\caption{Diagram of the superimposed constellation with IM operation for a two users case, where a far user is paired with a near user to perform NOMA.}
	\label{fig_3-Superimposed-Cons}
\end{figure*}

In this case, the symbol corresponding to the near user remains original state when index bit is $\left[ 0 \right] $ and rotate $\pi /2$ when the input index bits is  $\left[ 1 \right] $. To explain the superimposed symbols after the IM operations more clearly, the superimposed constellation for two users case is given in Fig. \ref{fig_3-Superimposed-Cons}. As shown in Fig. \ref{fig_3-Superimposed-Cons}, the minimum Euclidean distance for far user maintains the same as the original constellations, while the near user increases a double SE enhancement at a cost of a minor minimum Euclidean distance reduction after the rotation of constellation. In the low complexity SIC detector, the symbols of far users are detected first which will be subtracted before the detection of the near users, therefore the error probability of the symbols corresponding to the far users has greater impact on the system BER performance. For the near user in IM-NOMA-RC, the SIC detector not only needs detection the transmitted symbol, but also needs to calculate the rotation angel in IM operation to retrieve the index bits. Moreover, IM operation is performed on the symbol transmitted by the near users, which means 2 bits can be transmitted per subcarrier.

\section{IM-NOMA-RC Detection and Performance Analysis}\label{Detectors}
In IM-NOMA-RC system, the detection processes need to detect symbols of both its own and other users, which leads to more complicated detection procedures. In the following, the optimal ML detector and a low complexity SIC detector are considered to the detection of the transmitted symbols for each user.

\subsection{Maximum Likelihood Detection}
For the detection of the received signals at the $ n $th user, the optimal ML detector detects all the transmitted symbols by performing an exhaustive search of all possible superimposed symbols subcarrier to subcarrier, and then extracts its own information bits. Based on the received signal model given in the formula  (\ref{y-signal-n}), the detection of the $k$th subcarrier at the $ n $th user based on the ML criterion is given by
\begin{equation}\label{ML-Dection}
	\hat{x}_{n,k}=\text{arg}\underset{x\in \left\{ s,\varphi \right\}}{\min}\lVert y_{n,k}-h_{n,k}x \rVert ^2,
\end{equation}
where $x\in \left\{ s,\varphi \right\} $ denotes all the realizations of the superimposed symbols with different modulated symbols in conventional constellation and rotation of constellation in IM operation. 
According to the above detection processes, the computational complexity in terms of FLOPs of the ML detector per subcarrier is given by
\begin{equation}\label{FLOPs-ML}
	\eta _{ML}=3M^NC_{IM},
\end{equation}
where $C_{IM}=2^{p_2}$ denotes the number of realizations of rotation constellation in IM operations. 
Since the exhaustive search over all the realizations leads to a prohibitive computational complexity in the detection process, which makes the optimal ML detector can not be used in the realistic communication applications. Therefore, it is of great value to design an effective for IM-NOMA-RC while maintaining lower detection complexity. To this end, a serial of low complexity detection algorithms for IM-NOMA-RC system are proposed in next subsections.

\subsection{Low Complexity SIC Detector}
In the NOMA systems, SIC theory based detection algorithm is widely used in the detections due to its low computational complexity and variable number of the symbols in the detection process for different users. 
In SIC detection, each user needs to detect and subtract the symbols of the farther users before the detection of its own signal, which means that the $ n $th user needs to detect symbols with higher transmit power and applying interference cancellation operations. 
The same detection processes are repeated by the $ n $th user until its own symbols are obtained. 

In SIC detection, each ceases the detection processes when its own symbol is detected. Since only the near users employ IM operations in transmission, the detection processes have different calculation procedures. For the far users without IM operations, the detection is similar to conventional PD-NOMA system, which only needs to detect and cancel the father user's symbol. For the user $ U_{n} $ with $n\in \left\{ \text{1,2,}\dots ,B \right\} $ in the far user group, their transmitted symbols all drawn from conventional constellation. According to the principle of the SIC detector, the detection processes of the symbol corresponding to the $ l $th user by the $ n $th user be expressed as 
\begin{equation}\label{SIC-far-user}
	\hat{s}_{n,k}^{l}=\text{arg}\underset{s\in \left\{ \mathcal{S} \right\}}{\min}\lVert y_{n,k}^{l}-\alpha_{l}h_{n,k}s \rVert ^2,
\end{equation}
where $l=\text{1,2,}.\dots ,n$ denotes the index of user in the current detecting processes, $y_{n,k}^{l}=y_{n,k}^{l-1}-\alpha _{l-1}\hat{s}_{n,k}^{l-1}$ denotes the received signal after cancelling the farther user's signals, $ y_{n,k}^{1}=y_{n,k} $ denotes the original received signal by the $ n $th user. When the detection processes reach to itself, the users in far user group detects its own symbol and decode its own information bits. Then, the detection processes of the user $ U_{n} $ is finished. Note that, the detection in the IM-NOMA-RC system start from far users, which indicates that the farthest user $ U_{1} $ only needs to detect once to get its own symbol by treating the rest symbols from other users as noise and the user $ U_{n} $ needs to carry out $ n $ times of detection and cancelling calculations before its own signal is detected. Specifically, the nearest user $ U_{N} $ needs to repeat $ N $ times of the detection to get its own symbol.  

For the users in near user group, they need to detect all the symbols of far users, and cancel them in the received signal. For the near user $ U_{n} $ with $n\in \left\{ \text{B+1,B+2,}\dots ,N \right\} $ in the near user group, the detection processes when user $ U_{n} $ calculates the symbols corresponding to the far users are same to the detection in formula (\ref{SIC-far-user}). When calculating the symbols corresponding to the near users, the detection processes can be expressed as 
\begin{equation}\label{SIC-near-user}
\left< \hat{s}_{n,k}^{l},\hat{\theta} \right> =\text{arg}\underset{s\in \left\{ \mathcal{S},\theta \right\}}{\min}\lVert y_{n,k}^{l}-\alpha _lh_{n,k}e^{j\theta}s \rVert ^2 ,
\end{equation}
where $l=B+\text{1,}B+\text{2,}.\dots ,n$ denotes the index of user in the near user group, and $\theta $ denotes the rotation angel of the constellation in IM operations. In the detection, the user $ U_{n} $ needs to detection both modulated symbols for near users and determine the rotation angle to decode the index bits, $y_{n,k}^{l}=y_{n,k}^{l-1}-\alpha _{l-1}\hat{s}_{n,k}^{l-1}e^{j\hat{\theta}_{l-1}}$ denotes the received signal after cancelling the farther user's signals in near user group. 
Let us define $y_{n,k}^{B}=y_{n,k}^{B-1}-\alpha _{B-1}\hat{s}_{n,k}^{B-1}$ to denote the received signal when all the far users' symbols are subtracted from the original received signal. 
When calculation process reach $ l $th user's symbol, the received signal can be rewritten as $y_{n,k}^{l}=y_{n,k}^{B}-\sum_{i=B}^{l-1}{\alpha _i\hat{s}_{n,k}^{i}e^{j\hat{\theta}_i}}$. 
Compared with far users, near users can benefit from IM operations to transmit information bits in two ways. Therefore, more information bits can be transmitted with same orthogonal communication resource block for the near users. In addition, the index bits can also be information bits of another user in the near user group, which indicates that the proposed IM-NOMA-RC can support $ N+1 $ users within an orthogonal resource block. Let us define the special user whose information bits are transmitted by IM operations is $ U_{N+1} $. Note that, the user $ U_{N+1} $ need to detect all the symbols of other users as user $ U_{n} $, and has the most calculation complexity.

In IM-NOMA-RC, the IM operations are performed only on the near users, the users in far user grout and near user group have different detection processes. For the far users $U_n$ with $n\in \left\{ \text{1,2,}\dots ,B \right\} $, they detect symbols from conventional constellation and have same computational complexity as that of the classical PD-NOMA system. According to the above detection processes for both far users and near users, the computational complexity in terms of FLOPs of the SIC detector is summarized.  
Based on the detection processes given in formula (\ref{SIC-far-user}), the detection complexity in terms of FLOPs per subcarrier for far user group is given by
\begin{equation}\label{Flops-SIC-far}
	\eta _F=3Mn+n-1.
\end{equation}
For the users in the near user group, $U_n$ with $n\in \left\{ \text{B+1,B+2,}\dots ,N \right\} $ need to detect and subtract the farther users' signals first, and then detect both transmitted symbols and index bits from the rotated constellation. 
According to the detection processes, the detection complexity in terms of the FLOPs for the users in near user group can be expressed as
\begin{equation}\label{Flops-SIC-near}
	\eta _N=3MB+4MC_{IM}\left( n-B \right) +n-1.
\end{equation}
Compared with the optimal ML detector, the SIC detector, which decodes and cancels the signals of the farther users in one-by-one manner, can efficiently reduce the detection burden at the receiver side.

\subsection{Error Performance Analysis}
In this subsection, the average pairwise error probability (APEP) is derived in closed form for the ML detector by employing the union bounding technique. Since all the subcarriers follow the same detection procedures, we only present the analysis of the $ n $th user to illustrate the APEP of the proposed IM-NOMA-RC. The conditional pairwise error probability (PEP), which is defined as probability that the transmitted symbol $ x_{j} $ is erroneously detected as $x_{i}$ in the transmission, can be expressed as
\begin{equation}\label{CPEP-Q}
	p\left( \psi _{i,j,n}\left| h_n \right. \right) =Q\left( \sqrt{\frac{\psi _{i,j,n}}{2\sigma ^2}} \right) ,
\end{equation}
where $\psi _{i,j,n}=\lVert P_Th_n\delta _{i,j} \rVert ^2$, $\delta _{i,j}=\left( x_j-x_i \right) $, $ \sigma ^2 $ denotes the variance of AWGN noise. 
The unconditional PEP can be expressed as
\begin{equation}\label{UCPEP-Y}
	p\left( \psi _{i,j,n} \right) =P_r\left\{ \lVert y_n-P_Th_nx_j \rVert ^2>\lVert y_n-P_Th_nx_i \rVert ^2 \right\} ,
\end{equation}
By using the alternative formula of Q function, the unconditional PEP can be obtained by calculating the integration, which is given by
\begin{equation}\label{Integration-PEP}
	p\left( \psi _{i,j,n} \right) =\int_{-\infty}^{\infty}{p\left( \psi _{i,j.n}\left| h_n \right. \right)}f_{\psi _{i,j,n}}\left( \varphi \right) d\varphi ,
\end{equation}
where $ f_{\psi _{i,j,n}}\left( \varphi \right)  $ is the probability density function of $ \psi _{i,j,n} $. In transmission, the frequency channel response follows $h_n\sim \mathcal{C}\mathcal{N}\left( \text{0,}1 \right) $ Gaussian distribution, the random variable $ \psi _{i,j,n} $ still a zero mean Gaussian variable. Then, the APEP of the $ n $th user can be expressed as
\begin{equation}\label{APEP-x}
	P_e\le \frac{1}{p2^p}\sum_{x_i}{\sum_{x_j}{p\left( \psi _{i,j,n} \right) e\left( x_i,x_j \right)}},
\end{equation} 
where $ e\left( x_i,x_j \right) $ denotes the number of bits error for the corresponding pairwise error event.

\section{Numerical and Simulation Result}\label{Simulation-results}
In this section, the numerical simulation results of the proposed IM-NOMA-RC are provided and compared. In simulations, two users are served by the BS in the downlink transmission in the first case, where IM operation is performed only on the symbols transmitted to the near user.  The power coefficients for far user and near user in simulation is given by $\alpha _1=0.9$ and $\alpha _2=0.1$, respectively. In IM operation for near user, we define the rotation angle of the constellation is given by $\theta \in \left\{ \text{0,}\pi /2 \right\} $. In simulations, the system signal to noise ratio (SNR) is defined as the ratio of the total transmit power to the noise power. The subcarriers in each OFDM block is set as $L=128 $. For convenience of expression, we use ML and SIC to denote the detector employed in the simulations for both PD-NOMA and IM-NOMA-RC schemes. 

\begin{figure}[!t]
	\centering
	\includegraphics[width=3.6in]{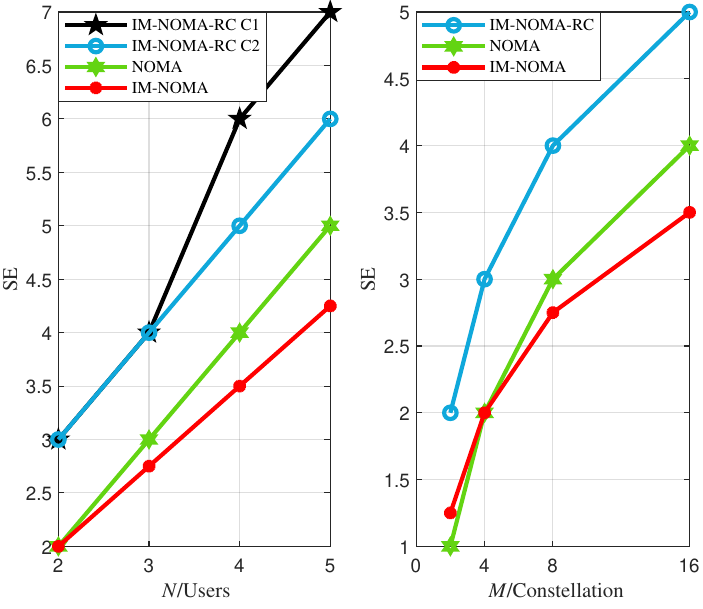}
	\caption{The SE comparison between NOMA, IM-NOMA-RC, and IM-NOMA.}
	\label{fig_SE-Comparesion}
\end{figure}

In Fig. \ref{fig_SE-Comparesion}, the SE of different NOMA schemes are compared, in which the first figure in Fig. \ref{fig_SE-Comparesion} is with different number of users and BPSK, and second figure in Fig. \ref{fig_SE-Comparesion} is with different order of constellations and two users. In first case of IM-NOMA-RC, which is denoted as IM-NOMA-RC C1, we assume that the system has more near users to perform IM operations to improve system SE, where $B= N-1$ users when $N={2,3}$, and $B=N-3$ users when $N=4,5$. While in the second case, which is denoted as IM-NOMA-RC C2, we set one user in near user group to perform IM operation. 
With enough users serve by a communication resource block, more users are utilized to perform IM can effectively improve the system SE as shown by the curve of IM-NOMA-RC C1. In IM-NOMA, each subblock has $ N_{S} =4$ subcarriers and $ K=3 $ activated subcarriers to transmit signals to multiple users.
Since there inactive subcarriers in IM-NOMA scheme, IM-NOMA has lower SE in most system parameter configurations. As shown in Fig. \ref{fig_SE-Comparesion}, IM-NOMA-RC has highest SE no matter what kinds of parameters is employed due to extra index bits transmitted by IM operations and no inactive subcarriers in transmission.

\begin{figure}[!t]
	\centering
	\includegraphics[width=3.6in]{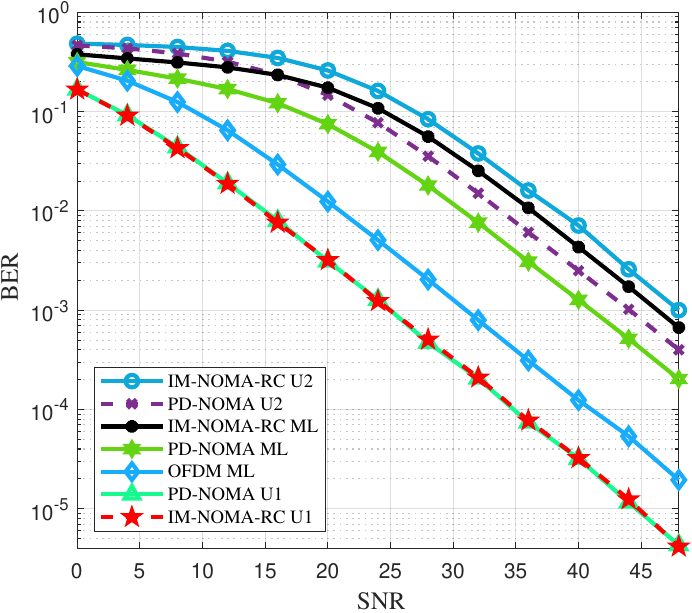}
	\caption{The BER comparison between NOMA, IM-NOMA-RC, and OFDM.}
	\label{fig_BER-All-scheme-Comparesion}
\end{figure}

In Fig. \ref{fig_BER-All-scheme-Comparesion}, the BER performance of proposed IM-NOMA-RC, NOMA and OFDM is compared with similar SE when the ML detector is employed. In IM-NOMA-RC, the system parameters are given by two users, and BPSK constellation. In this simulation, $ U_{1} $ and $ U_{2} $ are used to denote far user and near user for both IM-NOMA-RC and PD-NOMA, respectively. In PD-NOMA, also two user with BPSK constellation are employed. To achieve similar system SE, 8QAM constellation is employed in the OFDM system.  As expected, the IM operations have no influence on the far users and can transmit more information bits to the near users. As shown in Fig. \ref{fig_BER-All-scheme-Comparesion}, far use in both IM-NOMA-RC and PD-NOMA achieve almost the same BER performance, and outperforms OFDM in all the SNRs. For the near user, an error propagation in IM related detection leads to a BER performance loss. Note that, the near user in IM-NOMA-RC achieve double SE compared with that of near user in PD-NOMA. Although the BER performance of the optimal ML detector for PD-NOMA is superior than that of IM-NOMA-RC, PD-NOMA has lower SE with same constellation is employed.

\begin{figure}[!t]
	\centering
	\includegraphics[width=3.6in]{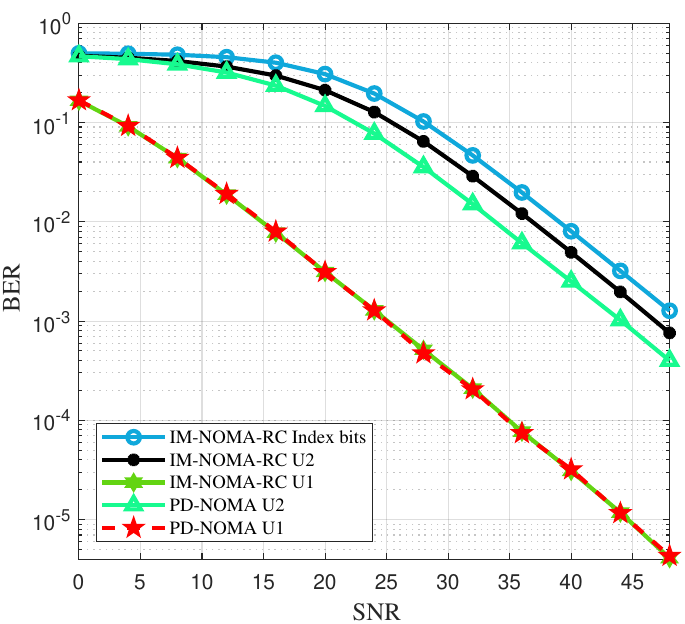}
	\caption{The BER comparison between far user, near user and index bits in IM-NOMA-RC.}
	\label{fig_BER-Three_Users}
\end{figure}

In Fig. \ref{fig_BER-Three_Users}, the BER performance of far user, near user without index bits and index bits are investigated with BPSK constellation. In this case, we assume that the information bits of the third user is only transmitted by IM implicitly. Since there are no modulated symbols transmitted to the third user, the power coefficients for the far user and near user are given by $\alpha _1=0.9$, $\alpha _2=0.1$, respectively. Similarly, we use $ U_{1} $, $ U_{2} $ and Index bits to distinguish the far user, near user and the third user in the simulation. From the numerical simulation results in Fig. \ref{fig_BER-Three_Users}, we can get the more insights of the BER performance of different users, where the far user with higher power level achieve better BER performance and index bits suffers the worse BER performance. This can be explained that the near user has lower power coefficient which diminishes the minimum Euclidean distance of the symbols in the superimposed constellation for the near user. In addition, IM can be an effective method to transmit information bits without allocating extra  power, whose information bits are transmitted by symbols of the near user implicitly. One more user can be served in IM-NOMA-RC even the transmit power of far user and near user maintains the same as PD-NOMA, which shows that IM can be effective ways to improve the system SE and serve more users.

\section{Conclusion}\label{Conclusions}
In this paper, a novel IM-NOMA-RC scheme is proposed for multiple users in downlink transmission. To perform rotation constellation based IM operations, the users in transmission are divided into two groups known as far user group and near user group. In IM-NOMA-RC, only the symbols for the near user group are performed IM operation to improve the system SE due to the extra bits transmitted by IM. As an independent dimension of modulation besides conventional modulated symbol modulation, IM operation can be used to transmit information bits for near users or serve one more user in transmission. Then, the optimal ML detector is studied for the IM-NOMA-RC system, and the upper bound on the BER performance by assuming the ML detector is derived in closed form. To reduce the detection complexity, the SIC theory based detector is proposed for both far users and near users.  In SIC theory based detections, the BER performance of far users maintains same as that of PD-NOMA system since the IM operation is performed on the near users. Finally, the numerical simulation results validate the superiority of the proposed IM-NOMA-RC scheme.

%\section{References Section}
%You can use a bibliography generated by BibTeX as a .bbl file.
% The IEEEtran BibTeX style support page is:
 
\bibliographystyle{IEEEtran}
\bibliography{mybibfileIMNOMA}

\newpage

\vfill

\end{document}